# Learning the Exact Topology of Undirected Consensus Networks

Saurav Talukdar[1], Deepjyoti Deka[2], Sandeep Attree[3], Donatello Materassi[4] and Murti Salapaka[3]

*Abstract*— In this article, we present a method to learn the interaction topology of a network of agents undergoing linear consensus updates in a non invasive manner. Our approach is based on multivariate Wiener filtering, which is known to recover spurious edges apart from the true edges in the topology. The main contribution of this work is to show that in the case of undirected consensus networks, all spurious links obtained using Wiener filtering can be identified using frequency response of the Wiener filters. Thus, the exact interaction topology of the agents is unveiled. The method presented requires time series measurements of the state of the agents and does not require any knowledge of link weights. To the best of our knowledge this is the first approach that provably reconstructs the structure of undirected consensus networks with correlated noise. We illustrate the effectiveness of the method developed through numerical simulations as well as experiments on a five node network of Raspberry Pis.

## I. INTRODUCTION

Consensus has emerged as a popular framework for multiple agents to interact and agree on a common value in a distributed manner. It has found applications in flocking of autonomous vehicles [1], resource apportioning in power systems to meet network demand [2], clock synchronization in wireless sensor networks [3], distributed optimization [4] and many more. Often the agents are modeled as nodes or vertices of a graph with the bi-directional interaction between a pair of agents being indicated as an undirected edge between them with specified weights.

In several consensus applications, the agents communicate over a wireless channel that is corrupted by additive receiver noise (modeled as a zero mean stochastic process) [5]. In this case, it is shown in [6] that the expected value of the agents' state converge to the average of the initial conditions asymptotically. Further, the deviation of the agent states from their mean functions is known to be a zero mean wide sense stationary (WSS) process. It is of interest to infer the underlying undirected graph of interaction between the agents from the measured time series of their states in a non-invasive manner (that is without actively perturbing the network). Such an algorithm is of interest for cyber security, where, the objective is to encrypt the relation between the users from an attacker. Other applications of this passive network structure learning approach is in fault detection and network augmentation for performance improvement [7], [8].

In this article we present an algorithm which can infer the undirected graph of interaction among the agents undergoing consensus dynamics from the measured time series of output of each agent. Notable works in the direction of topology learning from time series measurements of linear dynamical systems are [9], [10], [11], [12], [13], [14], [15], [16]. [11] and [12] adopt an active approach of inferring the topology by removing one node from the network at a time (known as node knockout). This may not be suitable from the perspective of an observer that cannot perturb the agents or the network structure due to operational or technical reasons. The approach presented in [13] uses graph lasso [17] to estimate the inverse of the steady state covariance matrix and then infers the topology. However, no analytical guarantees on the accuracy of the reconstruction are provided. [14], [15], [16] present topology learning algorithms for undirected consensus networks which utilize some knowledge of system parameters and also assume white noise model for the receiver noise. The framework presented here is not limited to a white noise model of receiver noise.

In this paper, we present a Wiener filtering based algorithm which provably learns the exact interaction topology of the agents from measured time series data. It is shown in [18] that Wiener filtering based topology learning of linear dynamical systems returns spurious links between nodes with a common child, in addition to the true links. In the present case of consensus over an undirected network, Wiener filtering based topology learning results in links being identified between all two hop neighbors. The principal contribution of this work is a method for detecting all spurious links identified by the Wiener filtering based reconstruction. We prove that Wiener filters associated with spurious links have a phase angle of $\pi$ over all frequencies based on which a two stage topology learning procedure is presented; in the first step we apply Wiener filtering to learn the topology, possibly with spurious links, following which we apply the pruning step of evaluating the frequency response of the Wiener filters to eliminate the spurious two hop neighbor links. Two notable features of this new algorithm are that, it recovers the exact topology even in the presence of loops or cycles, and does not use any knowledge of system parameters or noise fluctuations. We experimentally validate our learning framework on a test network of Raspberry Pis undergoing consensus dynamics.

In the next section we introduce notions from graph theory, which are utilized later, following which we introduce the framework of discrete time average consensus with additive noise. In Section IV the Wiener filtering based network topology reconstruction algorithm is discussed. Then we present the frequency response based test for elimination of spurious links as well as the two stage topology learning algorithm in Section V. In Section VI, simulation results and experimental demonstration of the algorithm are presented, followed by conclusions in Section VII.

[1]Saurav Talukdar is with Department of Mechanical Engineering, University of Minnesota, Minneapolis, USA, `sauravtalukdar@umn.edu`
[2]Deepjyoti Deka is with Theory Division, Los Alamos National Laboratory, Los Alamos, USA, `deepjyoti@lanl.gov`
[3]Sandeep Attree and Muti V. Salapaka are with Department of Electrical and Computer Engineering, University of Minnesota, Minneapolis, USA, `attre002@umn.edu, murtis@umn.edu`
[4]Donatello Materassi is with Department of Electrical Engineering and Computer Science, University of Tennessee, Knoxville, USA `dmateras@utk.edu`

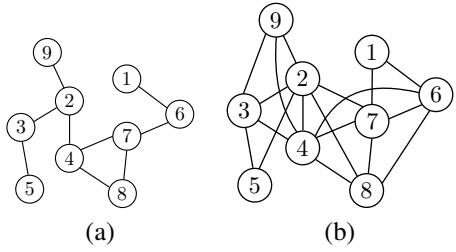

Fig. 1. (a) An undirected graph (nodes 2 and 3 are neighbors, 2 and 5 are two hop neighbors) and (b) its kin graph.

## II. PRELIMINARIES

In this section, basic notions of graph theory that are useful for the subsequent development are recalled [19].

*Definition 1 (Neighbors):* Given an undirected graph $\mathcal{G} = (\mathcal{V}, \mathcal{E})$, the set of neighbors of a node $i \in \mathcal{V}$ is, $\mathcal{N}_i := \{j : j \in \mathcal{V}, (i,j) \in \mathcal{E}\}$.

*Definition 2 (Two Hop Neighbor):* In an undirected graph $\mathcal{G} = (\mathcal{V}, \mathcal{E})$, the set of two hop neighbors of node $j$ is defined as, $\mathcal{N}_{j,2} = \{k | k \in \mathcal{V}, i \in \mathcal{V}, j \in \mathcal{V}$ such that $(j,i) \in \mathcal{E}$ and $(i,k) \in \mathcal{E}\}$. A node can be both a neighbor and a two hop neighbor.

*Definition 3 (Kin-graph):* Given an undirected graph $\mathcal{G} = (\mathcal{V}, \mathcal{E})$, its kin-graph is the undirected graph $\mathcal{G}_K = (\mathcal{V}, \mathcal{E}_K)$, where, $\mathcal{E}_K := \{(i,j) | i \in \mathcal{N}_j \cup \mathcal{N}_{j,2}, i \in \mathcal{V}, j \in V\}$. Figure 1 shows an undirected graph and its kin graph. The union of the set of neighbors and two hop neighbors of a node is referred to as its kin set.

Below we provide notations and properties pertaining to discrete time random processes.

The mean function $m(k)$ of a discrete time random process $x(k)$ is defined as $m(k) := \mathbb{E}[x(k)]$. The correlation function $R_x(k)$ of a wide sense stationary(WSS) random process $x(k)$ with zero mean ($m(k) = 0$ for all $k \in \mathbb{Z}$), is defined as $R_x(k) := \mathbb{E}[x(k)x(0)]$, while the cross correlation function $R_{x,y}(k)$ of jointly wide sense stationary(JWSS) random processes $x(k)$ and $y(k)$ with zero mean, is defined as $R_{x,y}(k) := \mathbb{E}[x(k)y(0)]$, where $\mathbb{E}[.]$ is the expectation operator. If $x(k)$ and $y(k)$ are uncorrelated, $R_{xy}(k) = 0$ for any $k \in \mathbb{Z}$.

*Definition 4:* (Power Spectral Density) The $z$ transform of the correlation function of a WSS random process $x(k)$ is called its power spectral density and is denoted as $\Phi_x(z)$. The cross power spectral density between two JWSS random processes $x(k)$ and $y(k)$ is the $z$ transform of the cross correlation function between them and is denoted as $\Phi_{xy}(z)$. If $x(k)$ and $y(k)$ are uncorrelated, $\Phi_{xy}(z) = 0$ almost surely.

## III. CONSENSUS DYNAMICS

Consider a collection of $m$ agents exchanging information with other agents, where, the interaction topology is given by an undirected graph $\mathcal{G} = (\mathcal{V}, \mathcal{E})$ with $\mathcal{V} = \{1, 2, ..., m\}$ being representative of the agents and $(j, i) \in \mathcal{E}$ with $i, j \in \mathcal{V}$, denote the edge between $i$ and $j$, that is, agent $j$ receives information from agent $i$ and vice versa. In the undirected graph $\mathcal{G}$, the neighbors of node $j$ are defined as the elements of its neighborhood set, $\mathcal{N}_j := \{i \in \mathcal{V} : (i,j) \in \mathcal{E}\}$. The two hop neighbors of node $j$ is defined as the elements of the set, $\mathcal{N}_{j,2} := \{i \in \mathcal{V} : (j,k), (i,k) \in \mathcal{E}, \text{for some } k \in \mathcal{V}\}$.

Let the state of agent $j$ be denoted as $x_j$. Agent $j$ updates its state according to the following rule,

$$x_j(k+1) = a_{jj}x_j(k) + \sum_{i \in N_j} a_{ji}x_i(k) + p_j(k) \quad (1)$$

where, $p_j(k)$ is the receiver noise of agent $j$ at instant $k$. The undirected nature of the interaction topology implies that, $a_{ji} \neq 0$ if and only if $a_{ij} \neq 0$. We assume that $a_{jj} \neq 0$, that is, each agent uses its past value to update its state. Also, all non zero entries of the matrix $A = [a_{ij}]$ are assumed to be non negative and less than 1. In vector form the network dynamics can be written as,

$$X(k+1) = AX(k) + P(k), \quad (2)$$

where, $X(k) = [x_1(k), x_2(k), \cdots, x_m(k)]^T$, $P(k) = [p_1(k), p_2(k), \cdots, p_m(k)]^T$ and the matrix $A \in \mathbb{R}^m \times \mathbb{R}^m$ is such that $A(j,i) = a_{ji}$. Each element of $P(k)$ is assumed to be zero mean WSS process uncorrelated with any other component of $P(k)$. Hence, the power spectral density matrix $\Phi_P(z)$ of $P(k)$, is a diagonal matrix and $\Phi_P(e^{\hat{j}\omega})$ is real and even function of frequency $\omega$. Furthermore, we assume the noise spectrum at any node $j$, $\Phi_{p_j}(z)$ to be positive almost surely (referred to as **topological detectability** condition). Note that $p_j(k), j = 1, ..., m$ can be any WSS process with self temporal correlations and not necessarily white noise. In the absence of receiver noise and $A$ being a doubly stochastic matrix with the underlying undirected graph being connected, it is known that the agents converge to the average of the initial conditions [20], that is,

$$\lim_{k \to \infty} x_j(k) = \frac{\sum_{i=1}^{m} x_i(0)}{m}, \quad (3)$$

where, $\{x_i(0)\}_{i=1}^{m}$ denote the initial values of the agents. In the presence of receiver noise the nodes do not converge to the average of the agent initial conditions. The mean of state vector $X(k)$ evolves according to the following dynamics [6],

$$\mathbb{E}(X(k+1)) = A\mathbb{E}(X(k)). \quad (4)$$

Consider, the deviation from the mean at each time step, that is,

$$\tilde{x}_j(k) := x_j(k) - \mathbb{E}(x_j(k)), j \in V, k = \{0, 1, ...\}$$

It can be shown that,

$$\tilde{x}_j(k+1) = \sum_{j \in \mathcal{N}_j \cup j} a_{ji}\tilde{x}_i(k) + p_j(k). \quad (5)$$

In compact form, the dynamics in terms of the deviation variables can be written as,

$$\tilde{X}(k+1) = A\tilde{X}(k) + P(k),$$

where, $\tilde{X}(k) := [\tilde{x}_1(k),...,\tilde{x}_m(k)]^T$. $\{\tilde{x}_j(k)\}_{j=1}^m$ is a collection of JWSS process as it is the output of linear filtering of uncorrelated WSS processes[21]. We present a topology learning algorithm which takes as input the time series $\{\tilde{x}_j(k)\}_{j\in\mathcal{V}}$ and outputs the interaction topology $\mathcal{G}$.

The discrete time dynamics of the deviation variable $\tilde{x}_j$ of the $j^{th}$ node in the $z$ domain can be cast as,

$$\tilde{x}_j(z) = \sum_{i\in\mathcal{N}_j} \mathcal{H}_{ji}(z)\tilde{x}_i(z) + e_j(z), j=1,2,\cdots,m. \quad (6)$$

where, $\mathcal{H}_{ji}(z) = \frac{a_{ji}}{S_j(z)}$, $e_j(z) = \frac{1}{S_j(z)}p_j(z)$ with $S_j(z) := z - a_{jj}$. Based on the assumptions on $P(k)$, $E(k) := [e_1(k),\cdots,e_m(k)]^T$ is a collection of uncorrelated zero mean vector WSS process, implying, the power spectral density matrix, $\Phi_E(z)$ is a diagonal matrix. Note that the diagonal entries of the matrix $H(z)$, $\mathcal{H}_{jj}(z) = 0$. Moreover if $i$ is not a neighbor of $j$ then $\mathcal{H}_{ji}(z) = 0, \mathcal{H}_{ij}(z) = 0$. In compact form the dynamics of (6) is written as,

$$\tilde{X}(z) = H(z)\tilde{X}(z) + E(z)$$

Thus the pair $(H(z), E(z))$ specifies the topology and network dynamics of the deviation variables. We further assume that $I - H(z)$ is invertible almost surely (referred to as **well posedness** condition).

## IV. Learning Topology using Wiener Filtering

We first present the concept of a non causal multivariate Wiener filter.

Let $v$ and $x_1,...,x_m$ be a collection of JWSS stochastic processes. Define $X(k) := [x_1(k),....,x_m(k)]^T$ and $\mathcal{X} := span\{x_1(k),...,x_m(k)\}_{k=-\infty}^\infty$. Consider the following least square optimization problem

$$\hat{v}(k) = \arg\inf_{q\in\mathcal{X}} \mathbb{E}(v(k) - q)^2. \quad (7)$$

If $\Phi_X(z) \succ 0$ (positive definite) almost surely, then the optimal solution $\hat{v}(k) \in \mathcal{X}$ exists, is unique (known as *multivariate non causal Wiener filter*) and is given by

$$\hat{v}(z) = \mathbf{W}(z)X(z), \mathbf{W}(z) = \Phi_{vX}(z)\Phi_X(z)^{-1}.$$

Refer [18] for further details. The next result details the properties of Wiener filter for nodal states of topologically detectable and well posed undirected consensus networks.

*Theorem 4.1:* Consider undirected graph $\mathcal{G}$ with $(H(z), E(z))$ specifying the network dynamics in (eqn:zdomnode) where nodal states are given by $\tilde{X}(k) = [\tilde{x}_1(k),...,\tilde{x}_m(k)]^T$. Define the space $\mathcal{X}_{\bar{j}} = span\{\tilde{x}_i(k)\}_{i\neq j, k=-\infty}^{k=\infty}$. The non-causal Multivariate Wiener filter $\hat{\tilde{x}}_j(k) \in \mathcal{X}_{\bar{j}}$ of the signal $\tilde{x}_j(k)$ is given by

$$\hat{\tilde{x}}_j(k) = \arg\min_{q\in\mathcal{X}_{\bar{j}}} \mathbb{E}(\tilde{x}_j(k) - q)^2. \quad (8)$$

An unique optimal solution to the above optimization problem $\hat{\tilde{x}}_j(k)$ exists and is given by

$$\hat{\tilde{x}}_j(z) = \sum_{i\neq j} W_{ji}(z)\tilde{x}_i(z) \quad (9)$$

where, $W_{ji}(z) \neq 0$ implies $i \in \mathcal{N}_j \cup \mathcal{N}_{j,2}$.

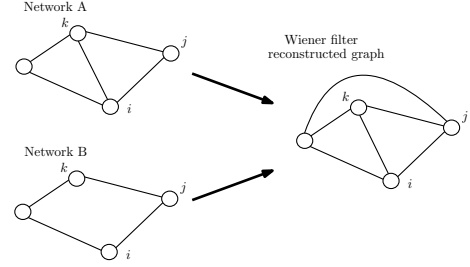

Fig. 2. Example of two undirected networks with consensus dynamics between the nodes, which result in the same kin graph obtained using multivariate Wiener filtering.

The proof follows from the main result of [18].

*Remark 1:* The above result does not guarantee that if $i \in \mathcal{N}_j \cup \mathcal{N}_{j,2}$, then $W_{ji}(z) \neq 0$. However, such cases are pathological (see [18]).

Thus the set of neighbors and two-hop neighbors of each agent in the network can be identified using non-zero Wiener filters and a kin graph $\mathcal{G}_K$ can be constructed. It is important to mention that the elimination of spurious edges in $\mathcal{G}_K$ is not obvious for undirected graphs with cycles. Consider the simple example in Fig. 2, where both the undirected cyclic graphs $A$ and $B$ result in the same reconstructed graph using multivariate Wiener filtering. It is difficult to determine if the Wiener filtering based reconstructed edge between $i$ and $k$ in Fig. 2 represents a true edge or a spurious edge between two hop neighbors. For radial networks (undirected connected networks with no cycles), it is possible to distinguish between true and spurious two hop neighbor edges due to a local topological separability rule as shown in [22], [23] or thresholding for static networks [24]. However for undirected graphs with cycles such topological separability results do not hold in general [25]. In the next section, a pruning step is presented, which eliminates the spurious two hop neighbor links obtained from Algorithm 1, and returns the true undirected graph from which the time series is measured.

## V. Pruning Two-hop Neighbors

Our pruning step is based on the phase response of Wiener filters. We first characterize the contribution of neighbors $\mathcal{N}_j$ and two-hop neighbors $\mathcal{N}_{j,2}$ of node $j$ to the multivariate non causal Wiener filter $\{W_{ji}(z)\}_{i=1,i\neq j}^m$.

*Theorem 5.1:* Consider undirected graph $\mathcal{G}$ with $(H(z), E(z))$ specifying the network dynamics in (6) where nodal states are given by $\tilde{X}(k) = [\tilde{x}_1(k),...,\tilde{x}_m(k)]^T$. The multivariate non causal Wiener filtering estimate $\hat{\tilde{x}}_j(k) = \sum_{i,i\neq j} W_{ji}(z)\tilde{x}_i(z)$ satisfies $W_{ji}(z) = \hat{N}_{ji}(z) + \hat{N}_{ji,2}(z)$, where,

$$\hat{C}_{ji}(z) = \frac{a_{ij}S_i(z)\Phi_{p_i}^{-1}(z)}{|S_j(z)|^2\Phi_{p_j}^{-1}(z) + \sum_{l\in\mathcal{N}_j} a_{lj}^2\Phi_{p_l}^{-1}(z)}, \quad (10)$$

$$\hat{P}_{ji}(z) = \frac{a_{ji}S_j^*(z)\Phi_{p_j}^{-1}(z)}{|S_j(z)|^2\Phi_{p_j}^{-1}(z) + \sum_{l\in\mathcal{N}_j} a_{lj}^2\Phi_{p_l}^{-1}(z)}, \quad (11)$$

$$\hat{N}_{ji}(z) = \hat{C}_{ji}(z) + \hat{P}_{ji}(z) \text{ and} \quad (12)$$

$$\hat{N}_{ji,2}(z) = -\frac{\sum_{k\in\mathcal{N}_j\cap\mathcal{N}_i} a_{kj}a_{ki}\Phi_{p_k}^{-1}(z)}{|S_j(z)|^2\Phi_{p_j}^{-1}(z) + \sum_{l\in\mathcal{N}_j} a_{lj}^2\Phi_{p_l}^{-1}(z)}. \quad (13)$$

*Proof:* It is shown in [18] that for nodes $i$ and $j$, the contribution $W_{ji}(z)$ is composed of two terms, $(\hat{C}_{ji}(z) + \hat{P}_{ji}(z))$ and $\hat{N}_{ji,2}(z)$, where,

$$\hat{C}_{j*}(z) = \frac{\Phi_{e_j}(z)H^*_{*j}(z)\Phi_e^{-1}(z)}{1+|H^*_{*j}(z)\Phi_e^{-1}H_{*j}(z)|\Phi_{e_j}(z)},$$

$$\hat{P}_{j*}(z) = (1 - \hat{C}_{j*}(z)H_{*j}(z))H_{j*}(z),$$

$$\hat{N}_{ji,2}(z) = -\hat{C}_{j*}(z)H_{*i}(z),$$

Here $H_{*j}(z)$ and $H_{j*}(z)$ denotes the $j-th$ column and $j-th$ row of the $H(z)$ matrix respectively and $H^*_{*j}(z)$ denotes complex conjugate transpose of the vector $H_{*j}(z)$. Substituting the expression for transfer functions in the context of an undirected network with linear consensus dynamics (see (6)) in the above equations, leads to the expressions described in the theorem statement. ∎

Note that $\hat{N}_{ji}(z)$ is the contribution to the Wiener filter for $i$ being a neighbor of $j$, while $\hat{N}_{ji,2}(z)$ is the contribution $i$ being a two-hop neighbor of $j$. We remark that it is possible for $i$ to be a neighbor and a two hop neighbor of $j$ simultaneously. Theorem 5.1 provides the machinery to determine the phase response of non-zero multivariate Wiener filter terms.

*Theorem 5.2:* In the undirected graph $\mathcal{G} = (\mathcal{V}, \mathcal{E})$, consider a well posed and topologically detectable consensus dynamics given by the pair $(H(z), E(z))$, with $\tilde{X}(k) = (\tilde{x}_1(k) \cdots \tilde{x}_m(k))^T$ as the output at time instant $k$. If $i$ and $j$ are strict two-hop neighbors in $\mathcal{G}$ such that $\mathcal{N}_i \cap \mathcal{N}_j \neq \{\}$ and $i \notin \mathcal{N}_j$, $j \notin \mathcal{N}_i$, then $\angle(W_{ji}(e^{\hat{j}\omega})) = -\pi$ for all $\omega \in [-\pi, \pi]$.

*Proof:* The proof directly follows from the expression of $\hat{N}_{ji,2}(e^{\hat{j}\omega})$ derived in Theorem 5.1. Clearly, $\hat{N}_{ji,2}(e^{\hat{j}\omega})$ is a negative number for $\omega \in [-\pi, \pi]$. ∎

The above theorem can identify spurious edges between nodes which are strict two hop neighbors if it does not hold for nodes that are true neighbors. The next theorem lists conditions under which $\angle(W_{ji}(e^{\hat{j}\omega})) = -\pi$ for all $\omega \in [-\pi, \pi]$ for nodes $i$ and $j$ that are neighbors.

*Theorem 5.3:* In the undirected graph $\mathcal{G} = (\mathcal{V}, \mathcal{E})$, consider a well posed and topologically detectable consensus dynamics given by the pair $(H(z), E(z))$, with $\tilde{X}(k) = (\tilde{x}_1(k) \cdots \tilde{x}_m(k))^T$ as the output at time instant $k$.

- If nodes $i$ and $j$ are such that $i \in \mathcal{N}_j$ and $i \notin \mathcal{N}_{j,2}$. Then, $\angle(W_{ji}(e^{\hat{j}\omega}))|_{\omega=0} = 0$.
- If $i$ and $j$ are such that $i \in \mathcal{N}_j$ and $i \in \mathcal{N}_{j,2}$ (both one and two hop neighbors), $\angle(W_{ji}(e^{\hat{j}\omega})) = -\pi$ for all $\omega \in [-\pi, \pi]$ if

$$Im(a_{ij}S_i(e^{\hat{j}\omega})\Phi^{-1}_{p_i}(e^{\hat{j}\omega}) + a_{ji}S^*_j(e^{\hat{j}\omega})\Phi^{-1}_{p_j}(e^{\hat{j}\omega})) = 0,$$

for all $\omega \in [-\pi, \pi]$, where, $Im(a + \hat{j}b) = b$ and $Re(a + \hat{j}b) = a$.

*Proof:*
- Since $i \notin \mathcal{N}_{j,2}$, $\hat{N}_{ji,2}(z) = 0$. It follows from (12) that,

$$W_{ji}(e^{\hat{j}\omega}) = N_{ji}(z),$$

where, it can be shown that the denominator of the expression on the right hand side is a positive real number for all $\omega \in [-\pi, \pi]$ while the numerator of the expression on the right hand side has a zero imaginary part at $\omega = 0$.

- $W_{ji}(e^{\hat{j}\omega}) = \hat{N}_{ji} + \hat{N}_{ji,2}$ where $\hat{N}_{ji}, \hat{N}_{ji,2}$ are given by (12) and (13). Both terms have the same denominator which is positive. Further the second term in the above equation has a zero imaginary part. Thus, for $\angle(W_{ji}(e^{\hat{j}\omega})) = -\pi$ for all $\omega \in [-\pi, \pi]$, imaginary part of the first term should be zero. The relation in the theorem ensures that. ∎

*Remark 2:* The consequence of Theorem 5.3 is that for nodes $i$ and $j$ that are neighbors but not two hop neighbors, $\angle(W_{ji}(e^{\hat{j}\omega})) \neq -\pi$ for all $\omega \in [-\pi, \pi]$. If $i$ and $j$ are true as well as two-hop neighbors, $\angle(W_{ji}(e^{\hat{j}\omega})) = -\pi$ for all $\omega \in [-\pi, \pi]$, only for unlikely cases because the system parameters have to take a specific set of values for the conditions in the theorem to be true at all frequencies. In other words, aside for pathological cases, the converse of Theorem 5.2 holds. We use this as a criteria to differentiate between true edges and spurious edges in the kin graph $\mathcal{G}_K$ (obtained by Wiener filtering) to recover $\mathcal{G}$.

*Learning Algorithm*

We now present Algorithm 1 that estimates the topology of an undirected graph $\mathcal{G}$ of agents undergoing consensus dynamics based on time-series of agent's state. The algorithm consists of two parts. The first part (Steps 1 - 9) determines the multivariate Wiener filter $W_{ji}(z)$ to estimate the output of each node $j$ from the rest of the time series. Edge set $\bar{\mathcal{E}}_K$ is populated by adding a link between each node pair $i, j$ if the $H_\infty$ norm of $W_{ji}(z)$ or $W_{ij}(z)$ is greater than a predefined threshold $\rho$ that indicates a non-trivial value (see Theorem 4.1). Thus $\bar{\mathcal{E}}_K$ estimates one and two hop neighbors. In the next part (Steps 10 - 15), we consider a finite set of frequency points $\Omega$ in the interval $[-\pi, \pi)$ and evaluate the phase angle of the Wiener filters corresponding to edges in $\bar{\mathcal{E}}_K$. If the phase angle is within a pre-defined threshold $\tau$ of $-\pi$, the algorithm designates them as spurious edges (see Theorem 5.2) and prunes them from $\bar{\mathcal{E}}_K$ to produce edge set $\bar{\mathcal{E}}$, which is an estimate of the edge set $\mathcal{E}$ of the true topology. In the limit of infinite data samples from each agent, $\bar{\mathcal{E}} \to \mathcal{E}$.

## VI. RESULTS

In this section, the performance of Algorithm 1 in estimating the consensus topology is demonstrated on a 5 node network depicted in Fig. 3(a). Note that in the test network, nodes $1, 2, 3$ are neighbors as well as two hop neighbors of each other. We present two set of results, first simulations using MATLAB software, and second experimental results on a network of Raspberry Pis.

For our first set of simulations, the receiver noise at each node is considered to be white noise that is generated using the 'wgn' function in MATLAB. The deviation variable time series is obtained from the nodal measurement time series by using the 'detrend'function in MATLAB. The reconstructed topology using Wiener filtering with $10^7$ samples from each node is shown in Fig. 3(b). The dashed

## Algorithm 1 Exact Topology Learning using Wiener Filtering

**Input:** agent deviation variable time series $\tilde{x}_i(k)$ for nodes $i \in \{1, 2, ..., m\}$ in $\mathcal{G}$, thresholds $\rho, \tau$, frequency points $\Omega$
**Output:** Estimate of True Edge Set $\bar{\mathcal{E}}$

1: **for all** $j \in \{1, 2, ..., m\}$ **do**
2:     Compute Wiener filter $W_j(z) = [W_{j1}(z) \cdots W_{jm}(z)]$ (see Appendix)
3: **end for**
4: Edge set $\bar{\mathcal{E}}_K \leftarrow \{\}$
5: **for all** $i, j \in \{1, 2, ..., m\}, i \neq j$ **do**
6:     **if** $\|W_{ji}(z)\| > \rho$ **then**
7:         $\bar{\mathcal{E}}_K \leftarrow \bar{\mathcal{E}}_K \cup \{(i, j)\}$
8:     **end if**
9: **end for**
10: Edge set $\bar{\mathcal{E}} \leftarrow \bar{\mathcal{E}}_K$
11: **for all** $i, j \in \{1, 2, ..., m\}, i \neq j$ **do**
12:     **if** $\pi - \tau \leq |\angle(W_{ji}(e^{j\omega}))| \leq \pi, \forall \omega \in \Omega$ **then**
13:         $\bar{\mathcal{E}} \leftarrow \bar{\mathcal{E}} - \{(i, j)\}$
14:     **end if**
15: **end for**

links are the spurious links recovered. Fig. 4(a) and Fig. 4(b) show the frequency response of Wiener filters between node 2 and all other nodes in Fig. 3(a) that are derived using Algorithm 1 with $10^7$ samples for each node. It is seen in Fig. 4(a) that the $H_\infty$ norm of $W_{25}(z)$ is separated from that of $W_{21}(z), W_{23}(z), W_{24}(z), W_{25}(z)$ which indicates that there exist no edge between nodes 2 and 5. Thus, the set of neighbors and two hop neighbors of node 2 can be recovered by choosing a proper threshold. Using the pruning step, the absolute values of the phase response of $W_{21}(z), W_{23}(z), W_{24}(z)$ are analyzed as shown in Fig. 4(b). It is clear that there is no edge between 2 and 4 (phase response being close to $\pi$) in the actual network and that they are two hop neighbors.

Fig. 5(a) shows the experimental setup, which consists of 5 Raspberry Pi [26] units that interact according to consensus dynamics with the interaction topology being the undirected graph of Fig. 3(a). The details of the experimental platform can be found in [27]. The communication channel among the units is through wifi and hence, packet drops do happen. The error (false negatives and false positives) performance of Algorithm 1 with respect to number of samples for simulations as well as experiments is shown in Fig. 5(b). It is seen that as the number of samples per node increases, the error decreases.

## VII. CONCLUSION

This article presented an algorithm which provably recovers the exact interaction topology of a group of agents undergoing consensus dynamics with ambient noise even in the presence of loops. The error in reconstruction decreases as more and more samples are used for learning the interaction topology. In the limit of large number of samples, the presented algorithm recovers the exact interaction topology of the agents. The effectiveness of the algorithm is demonstrated using MATLAB simulations as well as a network of Raspberry Pi devices. Learning the interaction topology is useful for follow-up analysis. In particular, we are interested in understanding the use of our learning framework in change detection in network topology, estimation of noise statistics as well as optimization of topology for improved convergence.

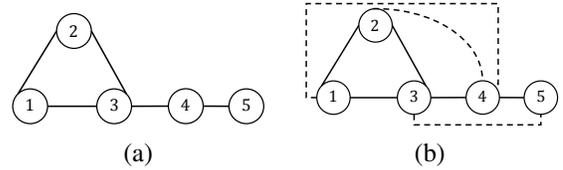

Fig. 3. (a) A 5 node undirected network with consensus dynamics between the nodes, (b) reconstructed topology of the 5 node network of Fig. 3(a) obtained using multivariate Wiener filtering with $10^7$ samples from each node. The dashed links are the spurious 2 hop neighbor links.

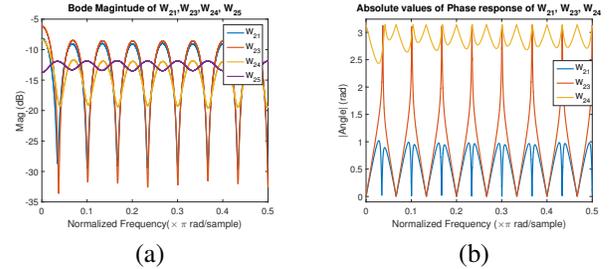

Fig. 4. (a) Bode magnitude plot of $W_{21}(z), W_{23}(z), W_{24}(z), W_{25}(z)$, (b) absolute values of phase response of $W_{21}(z), W_{23}(z), W_{24}(z), W_{25}(z)$. $W_{24}(z)$ has a phase response in the vicinity of $\pi$ for all frequencies, hence, is eliminated by the pruning step.

## VIII. ACKNOWLEDGMENTS

The authors S. Talukdar, D. Materassi and M. V. Salapaka acknowledge the support of ARPA-E for supporting this research through the project titled 'A Robust Distributed Framework for Flexible Power Grids' via grant no. DE-AR000071 and Xcel Energy's Renewable Development Fund. D. Deka acknowledges the support of funding from the U.S. Department of Energy's Office of Electricity as part of the DOE Grid Modernization Initiative.

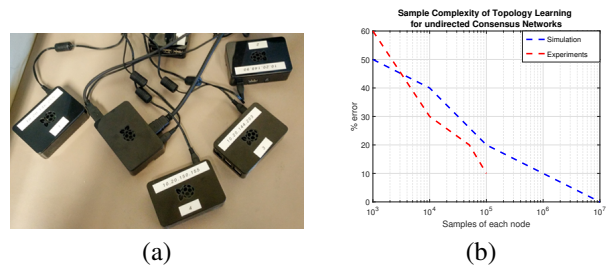

Fig. 5. (a) Experimental setup of 5 Raspberry Pi units interacting through wifi according to consensus dynamics with the interaction topology being the undirected graph in Fig. 3, (b) error percentage variation with number of samples per node in simulation as well as experiments.

## IX. APPENDIX: WIENER FILTER COMPUTATION[28]

The Wiener filter determines the optimal projection of a signal $x_j(n)$ in the space $\mathcal{X}_{\bar{j}} = span\{x_i(n+p) : p \in \mathbb{Z}\}_{i \neq j}$. Here we compute the Wiener filter by approximating it with a finite impulse response (FIR) filter, also known as FIR Wiener filter. Let the order of the FIR Wiener filter be $F$. Here the optimal estimate $\hat{x}_j(n)$ is written as,

$$\hat{x}_j(n) = \sum_{k \in \mathcal{V}, k \neq j} \sum_{p=-F}^{F} h_{k,l} x_k(n+l) \quad (14)$$

The Wiener filtering orthogonality condition is stated below and is used to determine the constants $h_{k,l}$ in $\hat{\theta}_j(n)$.

$$\mathbb{E}[\hat{x}_j(n) x_i(n+l)] = \mathbb{E}[x_j(n) x_i(n+l)], \quad (15)$$
for all $i \in \mathcal{V}, i \neq j, l \in \{-F, -F+1, \cdots, F-1, F\}$. \quad (16)

Thus, using (14) in (15),

$$[R_{\theta_1 \theta_i}(-F-l) \cdots R_{\theta_1 \theta_i}(F-l) \cdots R_{\theta_N \theta_i}(-F-l) \cdots R_{\theta_N \theta_i}(F-l)]h$$
$$= R_{\theta_j \theta_i}(-l), i \in \mathcal{V}, i \neq j, l \in \{-F, \cdots, F\}, \quad (17)$$
$$h := [h_1^T\ h_2^T \cdots h_{j-1}^T\ h_{j+1}^T \cdots h_N^T]^T, \text{and}$$
$$h_i^T := [h_{i,-F} \cdots h_{i,-1}\ h_{i,0}\ h_{i,1}\ \cdots\ h_{i,F}].$$

The set of equations in (17) describe $(2F+1)(N-1)$ linear equations in $(2F+1)(N-1)$ unknowns in the vector $h$. Thus, in combined form the equations in (17) can be written as

$$Rh = S$$

. Thus, $h = R^{-1}S$ is used to compute the coefficients of the Wiener filters. Note that the matrix $R$ and the vector $S$ can be computed using empirical correlations from the measured time series data. More details on numerical aspects of Wiener filtering can be found in [29], [30].